# Towards Rigorous Selection and Configuration of Cloud Services: Research Methodology

**Asmae Benali[1] and Bouchra El Asri[2]**

[1] IMS Team, ADMIR Lab, Mohammed V University, ENSIAS
Rabat, 8007/Agdal, Morocco

[2] IMS Team, ADMIR Lab, Mohammed V University, ENSIAS
Rabat, 8007/Agdal, Morocco

**Abstract**

Cloud computing has recently emerged as a major trend in distributed computing. We proposed a platform for selecting and configuring automatically an appropriate cloud environment that meets a set of consumer and provider requirements. It can easily adapt its behavior, either at design-time or runtime, to the change of the environment in matters of location, time, activity, interaction abilities, and communication restrictions. The platform based on the principles of dynamic software product lines (SPL), Agent-oriented software engineering, and the MAPE-k reference model. We based on the Design Science Research Methodology to conduct this work. In this article, we present the steps of our research following this methodology's guidelines.

***Keywords:*** *Cloud Computing, Dynamic Software Product Lines, Context-Aware System, Autonomic System, Design Science Research Methodology.*

## 1. Introduction

Cloud Computing is the on-demand access to shared computer system resources, namely data storage, applications, job processing, servers, networks, and the utilization of virtual machines [1]. It bases on utility computing to offer services, which is described as an information technology service, where the provider launches computing services and provides them on demand, following the "pay-as-you-go" model [2]. Therefore, the cloud provides several configuration choices to their customers in order to make the services customized. For that, the cloud developers have to deal with the variability of the cloud environment.

Nowadays, Context-Aware systems deployed on the cloud are becoming increasingly popular in many areas like mobile computing, adaptive and intelligent user interfaces robotics, etc.

Cloud computing is treated also as a runtime adaptative system which can adapt their services dynamically to both the cloud environment and consumer requirements change.

Furthermore, it has new challenges regarding the user context environment, such as location, activity, user preferences, mobile device constraints, resources, and communication capabilities. In fact, selecting and configuring continually a suitable cloud environment in both at design and run-time induces errors or complex configuration results which are generally made in an ad hoc manner. In consequence, our approach is to try to create an adaptative and context-aware platform for automatically select an appropriate cloud environment and also to find its configuration that at the same time meets and satisfies the consumer requirements and maximizing the provider profit.

Our platform bases on Software Product Lines (SPL) which is a widely used technology for reemploying the basic product components and structures by following planned variabilities models to derive various product families [3]. To manage the variable products and sous-products, SPL represents variability through variation points and variants. To ensure the reliability of these selection and configuration processes, we adapted dynamic software product lines which is a method that moves the product line engineering process to the run-time phase.

On the other side, we based on agents-oriented software engineering and the MAPE-k model to customize the selection and configuration cloud environment process regarding the consumer requirements and its environmental context.

To set up this model and answer the defined research questions, we needed to follow a rigorous research methodology such as Design Science Research Methodology (DSRM). Design Science is a paradigm that helps to provide answers to specific business issues (by implementing efficient solutions) [4] and the DSRM [5] is a systematic methodology based on this paradigm, which comes in a six steps process, starting from identifying research motivation, and up to evaluating the developed artifact. In this article, we present our research work and its results as conducted according to DSRM.

## 2. Background and motivation

In this section, we present briefly cloud computing, the selection and configuration of cloud services in literature.

### 2.1 Cloud computing

Cloud Computing is the provision of different services to its users, via the Internet, according to their requirements. These resources include tools and applications such as data storage, servers, databases, networks, and software. Before the birth of the term Cloud Computing, companies needed their proper computers or servers to their premises to use the software. In Cloud Computing, however, servers are hosted all over the world in data centers. Companies access their data by logging into their account. The number of cloud providers continues to grow which are companies that provide computer environments and services, namely software as a service (Software-as-A-Service -SaaS), platform as a service (Platform-As-A-ServIce -PaaS), and infrastructure as a service (Infrastructure-as-a-Service -IaaS). Many definitions for cloud computing have emerged in industry and academia. Cloud computing, for a majority of users, refers to a new IT paradigm with a source of savings. It serves to meet IT development needs readily by avoiding management problems.

The definition of Cloud Computing that is the most recognized and accepted by the scientific community is the one published by the National Institute of Standards and Technology (NIST) [1]. NIST defines cloud computing as an IT model that allows the network to have easy access to a shared set of configurable IT resources that can be quickly sourced and disseminated with minimal management or interaction effort with the service provider.

### 2.2 Selecting and Configuring Cloud Services

An application is deployed either on the IaaS of the cloud provider or on the PaaS that will run it to be accessible as a cloud service. In both cases, the application execution environment must be configured and scaled to take into account the heterogeneity and elasticity inherent in the cloud computing paradigm. There are indeed a large number of resources with heterogeneous levels of functionality and sizing possibilities. To deploy an application on a PaaS, you must first configure its environment, i.e., select a database for data storage, capacity, application server, CPU frequency required for execution, etc. Same for deployment on an IaaS that requires the configuration of the entire software stack (operating system, libraries, application servers) that supports the application and runs on the cloud provider's infrastructure. This variability in configuration and sizing offers several configuration possibilities, which are generally carried out on an ad hoc basis and are sources of errors when carried out by hand. In the next section, we look at literature approaches that address resource allocation in cloud computing.

## 3. An Application of Design Science Research Methodology for our Recherche Work

Information systems and the organizations in which they are used are complex, purposefully designed, and artificial. They are regrouped by people, structures, technologies, and work systems [6]. Hence, achieving rigor and relevance in information systems studies requires following a well-defined methodology during Research management. Two paradigms, in the literature, typify much of the research in the Information Systems discipline: behavioral science and design science [7]. Behavioral science aims to develop and verify theories that interpret or predict organizational or human behavior. Regarding design science, it aims to create new and innovative artifacts to solve organizational problems. Our research adopted the design science methodology to get the main purpose of our work, which is creating an optimal, dynamic, and context-aware platform for cloud services selection and configuration. That's why we have chosen this methodology; we focus more on the implemented system itself. In the next sections, we describe the design science process, its principal elements, and how it is instantiated in our research.

### 3.1 Design Science for a Rigorous Research Methodology

In our work, we followed the design science methodology for information systems. Design science is a set of methods, approaches, techniques, and perspectives (interpretative, critical, and complementary) to conduct research in Information Science [7]. Design science methodology is generally used to create and evaluate an artifact and/or a design theory whose purpose is to improve the current state of practice as well as existing research knowledge [8]. To structure our research project, we adopted the Design Science Research Methodology (DSRM) process model proposed by peffers [5]. This sequential process model consists of six main activities, namely: Identification and motivation of the problem, definition of solution objectives, design and development, demonstration, evaluation, and communication (see Fig. 1).

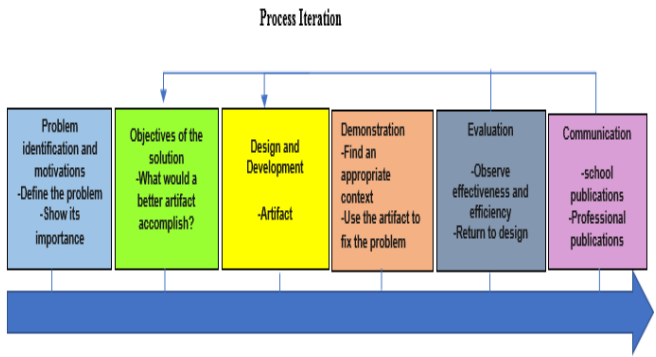

Fig. 1   Design science research methodology process

- Problem identification and motivation. This activity specifies the research problem and justifies the motivations of the research. It requires resources include the knowledge of the state of the problems, and the importance to fix it.

- Objectives of a solution. Deduce the objectives of a solution from the problem specification. They can be quantitative or qualitative. It requires resources include the knowledge of the state of the problems, and the existing solutions and their efficacy.

- Design and development. Design and develop the artifactual solution. the artifact can be constructs, models, methods, or instantiations. To move from the objectives to the design and development, it requires the knowledge of the theory which can be used to find a solution.

- Demonstration. After the development of the artifact, it needs to be demonstrated to verify its efficacy and efficiency to solve the problem. This could be through a simulation, a proof, a case study, experimentation. It is mandatory to know how to use the artifact to solve the problem.

- Evaluation. The goal of this step is to evaluate the created artifact. it observes and measures the effectiveness of a solution demonstrated to solve the problem defined. It can be based on some surveys result, quantitative performance measures, or a comparison study. This step requires knowledge of some metrics and analysis techniques.

- Communication. It aims to communicate the problem and its importance, the artifact, its benefit, and its effectiveness to researches, and technical audience for better comprehension and evaluation of the artifact [5].

3.2 Application of Design Science Research Methodology to our work

The fig. 2 depict how design science research methodology is applied to our study

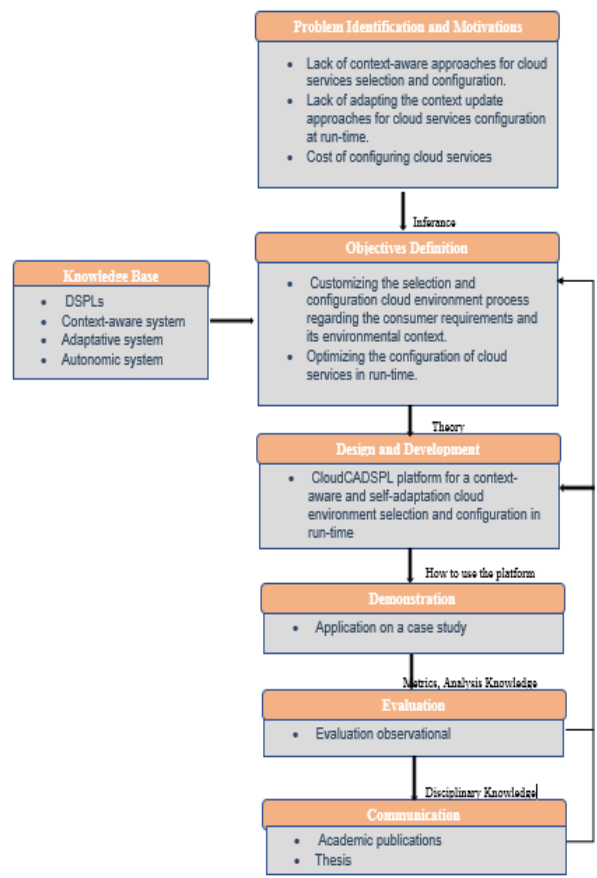

Fig. 2   DSRM applied to our research work.

- Problem Identification and Motivations

After a long search in the literature, we observe a lack of approaches that take into consideration the context-aware of the user to adapt dynamically the cloud services configuration in design and run-time.

- Objectives Definition

After identifying the problem, we start from the Knowledge base that contains other selection and configuration of cloud services approaches to address research challenges in a cloud computing environment. we put a list of research questions to determine the objectives of our work:

QR1: What is the knowledge base of this thesis?

QR2: How do we adapt the configuration of a cloud service to execution in order to meet a specific need?

QR3: How to express and manage the context variability of the cloud environment and that of the user in the cloud service configuration?

QR4: How to estimate this configuration to guarantee user satisfaction and maximize provider profit at the same time?

QR5: How to ensure the validity of this configuration and keep its traceability?

QR6: What are the means and tools to express, generate, implement, and verify the proposed solution? These research questions gave rise to a set of contributions that we give position in the next section.

- Demonstration

we applied our platform to a case study for the e-health system, precisely for e-diabetic system management.

- Evaluation

We evaluated our artifact using the same case study.

- Communication

We shared our research results by publishing papers in conferences proceedings and journals [10][11][12][13][14][15][16], which allow us to gather remarks to ameliorate our work.

## 4. Conclusions

Our research work aims to provide a platform based on the context-aware system and dynamic software product lines to automatically select an appropriate cloud service. This later should meet a set of requirements, define a configuration for the associated cloud environment and be able to adapt easily to the execution phase to the change of the environment either on the user side and/or cloud. In fact, the provider must offer resources that must be customizable to the different needs of the client and that meet their different requirements.

To conduct this work, we adopted the design science research methodology and followed its six steps to get coherent results. We defined the research problem, motivations, and the objectives of our research project. We suggested CloudCADSPL platform as an artifact and presented the results of our analysis. We applied our framework to an e-health case study.

In progress, we intend to evaluate our approach using metrics and thus we enhance our research methodology process.